\documentclass[aps,prl,twocolumn,groupedaddress,showpacs]{revtex4}
\usepackage{epsf}

\def\bea {\begin{eqnarray}}
\def\eea {\end{eqnarray}}
\def\be {\begin{equation}}
\def\ee {\end{equation}}
\def\ben{\begin{enumerate}}
\def\een{\end{enumerate}}
\def\bi{\begin{itemize}}
\def\ei{\end{itemize}}

\def\etal{{\it et al.}}
 \def\F{{\cal F}}
\def\prl {Phys. Rev. Lett.\ }
\def\pl {Phys. Lett.\ }
\def\pr {Phys. Rev.\ }
\def\np {Nucl. Phys.\ }
 \def\GV{G_{\mbox{\tiny V}}}
 
 \def\DRV{\Delta_{\mbox{\tiny R}}^{\mbox{\tiny V}}}
\def\hyphen{{\mbox{-}}}

\def\2p{|2p\rangle }
\def\4p2h{|4p\hyphen 2h\rangle }
\def\6p4h{|6p\hyphen 4h\rangle }

\begin{document} 
\preprint{ }
 
\title{Superallowed Beta Decay of Nuclei with $A \geq 62$:
\\The Limiting Effect of Weak Gamow-Teller Branches}

\author{J.C. Hardy}
\author{I.S. Towner}
\altaffiliation{Present address: 
Department of Physics,
Queen's University, Kingston, Ontario K7L 3N6, Canada}
\affiliation{Cyclotron Institute, Texas A \& M University,                    
College Station, Texas  77843}
\date{\today} 
\begin{abstract} 

The most precise value of $V_{ud}$, which is obtained from
superallowed nuclear $\beta$ decay, leads to a violation of CKM
unitarity by $2.2\sigma$.  Experiments are underway on two
continents to test and improve this result through decay studies
of odd-odd $N = Z$ nuclei with $A \geq 62$.  We show, in a series of
illustrative shell-model calculations, that numerous weak Gamow-Teller
branches are expected to compete with the superallowed branch
in each of these nuclei.  Though the total Gamow-Teller strength
is significant, many of the individual branches will be unobservably
weak.  Thus, new techniques must be developed if reliable $ft$-values
are to be obtained with 0.1\% precision for the superallowed branches.

\end{abstract} 


\maketitle

One of the most exacting tests of the unitarity of the Cabibbo-Kobayashi-Maskawa
(CKM) matrix is provided by nuclear $\beta$-decay. Precise measurements of the
$ft$-values for superallowed $\beta$-transitions between $T = 1$ analog $0^+$
states are used to determine $\GV$, the vector coupling constant; this, in turn,
yields $V_{ud}$, the up-down element of the CKM matrix.  In contradiction to the
Standard Model, the result from current world data violates CKM unitarity by more
than two standard deviations \cite{WEIN98}: {\it viz.} $V_{ud}^2 + V_{us}^2 +
V_{ub}^2 = 0.9968 \pm 0.0014$.  The potential significance of this outcome has
drawn attention to the reliability of small theoretical corrections that must be applied to each
experimental $ft$-value in order to extract $\GV$.  In particular, there are numerous
experimental programs now under way \cite{Ba01,Pi01,Oi01,He01,Hy99,Bl01,Ro01,Ga01,Fe01} to
study the superallowed decays of odd-odd $N = Z$ nuclei with $A \geq 62$, where
the charge-dependent correction terms are expected to be larger than among the
lower-$Z$ nuclei where all previous measurements have been made \cite{WEIN98,Ha90}.

Here, we report illustrative calculations demonstrating that these heavier nuclei will also exhibit a
serious complication that is not present among the lower-$Z$ nuclei.  This complication --
the presence of numerous weak Gamow-Teller $\beta$-decay branches that, in total, will
compete with the superallowed branch -- must ultimately limit the precision achievable
on any superallowed $ft$-value in this mass region.  It is particularly important to
recognize that many of these branches can be below the threshold for conventional
$\gamma$-ray detection and could easily be ignored.  If they are ignored, any $ft$-value
quoted for the superallowed branch could certainly not be relied upon to the $0.1\%$
precision required for a demanding test of the charge-dependent corrections.  Our
results have a similar impact on measurements of non-analog $0^+ \rightarrow 0^+$
$\beta$-transitions in nuclei with $A \geq 62$.  However, they do {\it not} indicate any
problems with the superallowed $ft$-values previously measured for lighter
($A \leq 54$) nuclei.  Thus, the non-unitarity result stands unaltered.

In general, $\GV$ can be obtained from the measured $ft$-value of a $0^+ \rightarrow 0^+$
$\beta$-transition between $T = 1$ analog states via the relationship \cite{WEIN98}
\begin{equation}
\F t \equiv ft (1 + \delta_R)(1 - \delta_C ) =  
\frac{K}{2 \GV^2 (1 + \DRV )} ,
\label{Ft}
\end{equation}
\noindent with
\begin{eqnarray}
K/(\hbar c)^6 & = & 2 \pi^3 \hbar \ln 2 / (m_e c^2)^5
\nonumber \\
& = & ( 8120.271 \pm 0.012) \times 10^{-10} {\rm GeV}^{-4}
{\rm s},
\label{K}
\end{eqnarray}
\noindent where {\it f} is the statistical rate function, {\it t} is the partial half-life
for the transition, $\delta_C$ is the isospin-symmetry-breaking correction, $\delta_R$ is
the transition-dependent part of the radiative correction and $\DRV$ is the transition-independent
part.  Here we have also defined $\F t$ as the ``corrected" {\it ft}-value.  Even though
the three calculated correction terms, $\DRV$, $\delta_R$ and $\delta_C$, are all of order
1\%, their estimated uncertainties ($\sim 0.1\%$) actually dominate the uncertainty in the derived value
of $\GV$.  Thus, improvements in the unitarity test must be sought through improvements in
the precision of these calculations rather than improvements in the experimental input.

Because the leading terms in the radiative corrections are well founded in QED \cite{WEIN98},
attention has focused more on the isospin symmetry-breaking correction, $\delta_C$.  It is
somewhat smaller than the radiative corrections but depends strongly
on the structure of the nuclear states involved.  (Actually, a small component of the
$\delta_R$ term also depends on nuclear structure but, for our present purposes, that
component need not be explicitly identified.)  The term $\delta_C$ comes about because
the Coulomb and charge-dependent nuclear forces break isospin symmetry between the
analog initial and final states in superallowed $\beta$ decay.  Although small, this term is
clearly very important: for the nine precisely-measured superallowed transitions in nuclei
between $^{10}$C and $^{54}$Co, the transition-to-transition variations that appear in the
uncorrected experimental $ft$-values would not themselves pass the CVC test.  It is
only after the $\delta_C$ (and $\delta_R$) corrections have been applied, yielding the
corrected $\F t$-values, that the results are constant to three parts in $10^4$.  This
in itself can be considered some validation of the $\delta_C$ calculations, providing we
accept CVC in the first place, but a more precise test of their validity would be welcome.

The important role of the $\delta_C$ correction in the unitarity test has focused recent 
attention on odd-odd $T_z = 0$ nuclei with $A \geq 62$ since the calculated $\delta_C$ corrections for their superallowed $\beta$-decay transitions are computed to be
relatively large \cite{OB95,SGS96}.  The interest in these nuclei has also coincided
with the emergence of new radioactive-beam facilities, which for the first time
have made it possible, at least in principle, to produce them in statistically
significant quantities.  The expected values of $\delta_C$, as calculated by Ormand
and Brown \cite{OB95}, are in the 1.0 to 1.8\% range, depending on the nucleus in
question and on the calculational method adopted; that is, three to five times
larger than the $\delta_C$ values for the nine precisely measured cases with
$A \leq 54$.  The main reason for this dramatic increase is the predominance
of the $2p$ shell-model orbital in nuclei with $A \geq 62$.  The radial wave function
for this orbital has a node, while the important orbitals in the lighter nuclei
are all nodeless, and this has a strong impact on the matrix elements of the Coulomb interaction.
Because the range of calculated $\delta_C$ values would be considerably increased by the
inclusion of nuclei with $A \geq 62$, it is argued that the calculations could be tested
more stringently against CVC by precise $ft$-value measurements in that mass region.

To measure an {\it ft}-value, three quantities are required from experiment: the transition
energy, $Q_{EC}$, which is used in calculating {\it f} ; the half-life, $t_{1/2}$, of the parent
nuclide and the branching ratio for the superallowed transition, which together yield the
partial half-life, {\it t}.  Recent experimental activity on the $A \geq 62$ superallowed
emitters includes the $Q$-value \cite{He01}, half-life \cite{Ba01}, and branching ratios
\cite{Pi01,Oi01}for $^{74}$Rb decay; the branching ratios for $^{62}$Ga decay \cite{Hy99,Bl01,Ro01};
and the half-lives of heavier nuclei up to $^{98}$In \cite{Ga01,Fe01}.  Most of the results
from these measurements are not yet at the required 0.1\% level of precision, but the quality
is improving steadily.

The purpose of this letter is to point out an important -- and complicating -- property
of the decays of these $A \geq 62$ nuclei.  As with the lighter odd-odd $T_z = 0$
nuclei, we expect their ground-state-to-ground-state superallowed branch to be
predominant; however, unlike the lighter cases, that branch will not constitute
$\geq 99.94 \%$ of the total decay rate, but instead will amount to around $99.0 \%$  for
$62 \leq A \leq 74$, and somewhat less for the heavier nuclei.  What makes
this difference critical is that the remaining $\sim 1\%$ $\beta$-decay strength
is expected to be spread over numerous Gamow-Teller transitions, of which all
those stronger than, say, $0.01 \%$ will have to be identified and measured in
order to determine the superallowed branching ratio to the required $0.1 \%$
precision.  The existance of these Gamow-Teller branches simply follows from
the fact that, as one moves to heavier and heavier $T_z = 0$ nuclei in the same
$A = 4n + 2$ sequence, the $\beta$-decay $Q$-value increases, thus opening up a larger
and larger energy window for $\beta$ decay.  At the same time, the density of $1^+$
states in the daughter also increases, as does their structural complexity, with the
result that weak Gamow-Teller branches become abundant.  The deleterious effects of
numerous weak Gamow-Teller transitions have been remarked in the study of much heavier
exotic nuclei \cite{JH7784} but their potential impact on precise superallowed
$ft$-values has not been noted before.

To quantify these ideas, and to illustrate the nature of the problem, we have
mounted a series of shell-model calculations for decays of the four $4n + 2$
nuclei with $62 \leq A \leq 74$ and, for comparison purposes, the three cases
with $46 \leq A \leq 54$, where precise data already exist.  Since these are
just illustrative calculations, the model spaces were kept fairly modest.  For
$46 \leq A \leq 54$, we took a $^{40}$Ca core with a
$(f_{7/2})^{n-r} (p_{3/2},f_{5/2},p_{1/2})^r$
model space truncated to $r \leq 3$.  We used standard effective interactions,
KB3 \cite{KB66,PZ81} and FPM13 \cite{Ri91}, but, because of the truncations
we readjusted their centroids to reproduce the experimental splitting between
the ground-state $0^+$ and the first-excited $0^+$ state, a key datum
for superallowed beta decay.

For nuclei with $62 \leq A \leq 74$, we use the model space
$(p_{3/2},f_{5/2},p_{1/2})^n$,
which is built on a $^{56}$Ni core with an effective interaction from Koops and
Glaudemans \cite{KG77} based on the modified surface-delta interaction (MSDI).  For
$A = 62$, 66 and 70, this interaction puts the excited $0^+$ close to its observed
location, but in $^{74}$Kr it fails badly, placing the state at 2.5 MeV, compared
to the experimentally known \cite{Be99} 0.5 MeV excitation.  Thus, for $A = 74$, it
is essential to include configurations involving the $g_{9/2}$, $d_{5/2}$ and possibly
the $g_{7/2}$ orbitals.  Such calculations quickly become unmanageable, so we limited
the $d,g$-shell occupation to two nucleons and tuned the effective interaction
(MSDI$^{\prime}$) to reproduce the energy of the first-excited $0^+$ state.

Clearly, such simple calculations
cannot possibly be expected to reproduce the properties of the $62 \leq A \leq 74$
nuclei in detail.  It is well known that $^{74}$Kr is difficult to describe in the
shell model since it is in a region where deformation effects are growing.  Indeed
projected Hartree-Fock-Bogoliubov calculations \cite{Pe00} point to oblate-prolate
shape coexistence in the low-energy spectrum.  Furthermore, even in the lighter $A = 62$,
66 and 70 nuclei, the $g_{9/2}$ orbit is becoming important, particularly for high-spin
states as demonstrated by Vincent \etal \cite{Vi98} for $^{62}$Ga.  However, since our
simple calculations already demonstrate considerable complexity in the beta decay of all these
nuclei -- the principal thrust of this communication -- our conclusions will not be altered
by larger calculations with additional orbitals, which can only serve to increase
this complexity. 

In Table~\ref{t:branch} we present the results of these shell-model
calculations.  In the sixth column, we identify how many $1^+$ states
in the daughter $T_z = 1$ nuclei are calculated to have an
excitation energy, $E_x$, less than the electron-capture $Q$-value,
$Q_{EC}$.  For each of these $1^+$ states we computed the Gamow-Teller
transition probability, $B(GT)$, and the partial width, $\Gamma$, with
\be
\Gamma \propto B(GT)~f ,
\label{Width}
\ee
where $f = f_+ + f_{ec}$ is the sum of the statistical rate functions
for positron decay and electron capture.  The proportionality constant is
fixed from the ground-state-to-ground-state superallowed branch, with the
result that the Gamow-Teller branching ratios, $BR$, are given by
\be
BR = \frac{\Gamma}{\Gamma_0} = \frac{B(GT)~f}{2 f_0},
\label{BR}
\ee where $\Gamma_0$ and $f_0$ are the width and statistical rate
function for decay to the ground state.  We sum these branching ratios over
all the $1^+$ states in the $Q$-value window and present the results in
column seven.

{\small

\begin{table}[h]
\begin{center}
\caption{Summed Gamow-Teller branching fractions in the superallowed
decay of selected $A = 4 n + 2$ nuclei.
\label{t:branch}}
\vskip 1mm
\begin{tabular}{lllllcl}
& & & & & & \\[-6mm]
\hline
\hline
\multicolumn{1}{c}{Parent} & &
\multicolumn{1}{c}{Shell} &
\multicolumn{2}{c}{First $1^+$state} &
\multicolumn{1}{c}{$\#$ of} &
\multicolumn{1}{c}{Total GT} \\
\cline{4-5}
\multicolumn{1}{c}{Nucleus} &
\multicolumn{1}{c}{$Q_{EC}$} &
\multicolumn{1}{c}{model} &
\multicolumn{1}{c}{Expt.} &
\multicolumn{1}{c}{Theo.} &
\multicolumn{1}{c}{$1^+$states\footnotemark[3]} &
\multicolumn{1}{c}{branching\footnotemark[3]} \\
&
\multicolumn{1}{c}{(MeV)} & &
\multicolumn{1}{c}{(MeV)} &
\multicolumn{1}{c}{(MeV)} & &
\multicolumn{1}{c}{($\%$)} \\
\hline
& & & & & & \\[-3mm]
$^{46}$V  & ~7.051 & FPMI3 & ~3.73 & 4.18 & ~~7 & ~~~~~0.027 \\
          &       & KB3   &      & 2.34 & ~10 & ~~~~~0.020 \\
$^{50}$Mn & ~7.632 & FPMI3 & ~3.63 & 3.91 & ~16 & ~~~~~0.013 \\
          &       & KB3   &      & 3.54 & ~35 & ~~~~~0.019 \\
$^{54}$Co & ~8.243 & FPMI3 & (3.84)\footnotemark[1] 
& 4.20 & ~23 & ~~~~~0.006 \\
          &       & KB3   &      & 4.17 & ~75 & ~~~~~0.024 \\
$^{62}$Ga & ~9.171 &  MSDI & (3.16)\footnotemark[1] 
& 2.48 & 110 & ~~~~~0.28 \\
$^{66}$As & ~9.57\footnotemark[2]  &  MSDI & (3.24)\footnotemark[1] 
& 2.27 & 255 & ~~~~~0.67 \\
$^{70}$Br & ~9.97\footnotemark[2]  &  MSDI & (3.14)\footnotemark[1] 
& 2.71 & 325 & ~~~~~1.59 \\
$^{74}$Rb & 10.418  &  MSDI & (3.2)\footnotemark[2] 
& 2.69 & 180 & ~~~~~0.72 \\
          &       & MSDI$^{\prime}$  &      & 2.76 & $>400$ & ~~~~~0.92 \\
\hline
\hline
\end{tabular}
\vspace{-3mm}
\footnotetext[1]{Lowest daughter state listed in ENSDF data file \cite{NDC} whose spin-parity
is unassigned.}
\footnotetext[2]{Assumed similar to the other $A \geq 62$ nuclei.}
\footnotetext[3]{These results have been derived by our shifting the
theoretical $1^+$ spectrum so that the energy of the lowest $1^+$ state
agrees with the experimental value or estimate.}
\end{center}
\end{table}

}

For the three well known cases with $46 \leq A \leq 54$, there are relatively few
Gamow-Teller transitions predicted and their total strength is a barely significant
$0.025 \%$ (or less) of the total $\beta$-decay.  This is in excellent qualitative
agreement with experiment.  All three nuclei can be produced prolifically and their
simple decays have been studied experimentally \cite{Ha94} with a sensitivity to branches
as small as 0.001\%. The total observed Gamow-Teller branching was 0.011\% (in one branch),
0.058\% (in two) and $\leq 0.001\%$ for the decays of $^{46}$V, $^{50}$Mn and $^{54}$Co
respectively.  These branches are already incorporated in the current analysis
\cite{WEIN98} of superallowed $\beta$-decay data.  Thus, our calculations offer no
correction whatsoever to those data.

\begin{figure}
\leavevmode
\epsfxsize=7cm
\hspace{-0.5cm}
\epsfbox{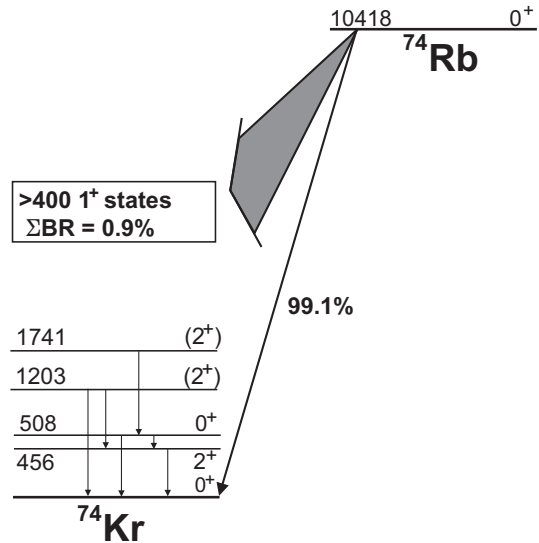}
\caption{Decay scheme of $^{74}$Rb.  All known low-spin states \cite{Pi01,NDC,Ru97} are shown
for $^{74}$Kr. The $1^{+}$ states and $\beta$-branching ratios are those calculated with MSDI$^{\prime}$.}
\label{fig:1}
\end{figure}

However, for the nuclei with $62 \leq A \leq 74$, where new measurements
are underway, our calculations indicate that the Gamow-Teller branching fraction
is substantially larger, ranging from $0.3 \%$ in $^{62}$Ga to $1.6 \%$
in $^{70}$Br, and certainly cannot be ignored.  Furthermore, it is also important to
recognize that these are accumulated branching fractions, the sum of many
individual branches.  The largest single (non-superallowed) branch calculated
in each case is about one-third of the total: $0.1 \%$ to the sixth $1^+$
state in $^{62}$Zn; $0.2 \%$ to the third $1^+$ state in $^{66}$Ge; $1 \%$
to the third $1^+$ state in $^{70}$Se; and $0.3\%$ to the fifth $1^+$ state
in $^{74}$Kr.  The remaining two-thirds of the Gamow-Teller strength in each decay is spread
over a large number of states: for example, in the case of $^{74}$Rb decay,
there are 20 transitions with individual branching ratios above $0.005\%$.  To date,
none of the $1^+$ daughter states has even been located and, in most cases, the
$\beta$-decay branches feeding them will be below normal detection sensitivity for
such exotic nuclei.  See Fig.~\ref{fig:1}.

There is clear experimental support for these predictions of complexity.  First, the
observed decays of $^{62}$Ga \cite{Hy99} and $^{74}$Rb \cite{Pi01} show evidence for the
population of states in their daughters that could not be fed directly by allowed $\beta$-decay
but must have been populated by unobserved $\gamma$ transitions from weakly fed states at higher
excitation.  Second, multiple Gamow-Teller transitions of the type we describe have been observed
\cite{NDC} in the decays of odd-odd $0^{+}$ nuclei with $N \neq Z$ in this mass region: $^{64}$Ga,
$^{66}$Ga and $^{78}$Rb.  All three exhibit complex decays with a minimum of 9, 12 and 23
significant Gamow-Teller transitions respectively, which populate $1^{+}$ states in their daughters.
All three have $Q_{EC}$ values that are {\it lower} than the $N=Z$ nuclei listed in
Table~\ref{t:branch} and the measured log$ft$-values are between 5 and 8, quite comparable to
those calculated for the latter.

For the superallowed emitters with $46 \leq A \leq 54$, the $1^+$ states populated in their
daughters de-excite by $\gamma$-ray (and conversion-electron) emission, so the
Gamow-Teller branching ratios are normally obtained experimentally from the intensities of observed
$\beta$-delayed $\gamma$-rays.  If most of the $\beta$-fed $1^+$ states
de-excite through the first excited $2^+$ state in the daughter, then the
$E2$ $\gamma$-ray from that $2^+$ state to the ground state might serve
as a ``collector", whose intensity approximates the total Gamow-Teller $\beta$
intensity. To examine this possibility, we have again used the shell model
to calculate all the de-excitation gamma-ray transition probabilities.
The results appear in Table~\ref{t:frac} where, for each
decay, we list the fraction of $\beta$-fed $1^+$ states that de-excite through
the first $2^+$ state.  Evidently, the excited $2^+$ state does act to some extent as a
collector but it misses enough of the Gamow-Teller strength that a measurement of
its intensity alone would not be sufficient for a precise quantitative determination of the
ground-state superallowed branching ratio.  
 
{\small

\begin{table}[h]
\begin{center}
\caption{Fraction of summed Gamow-Teller strength decaying through the         
lowest $2^+_1$ state.
\label{t:frac}}
\vskip 1mm
\begin{tabular}{lccc}
& & & \\[-4mm]
\hline
\hline
\multicolumn{1}{c}{Parent} &
\multicolumn{1}{c}{Shell} & 
\multicolumn{1}{c}{Summed GT} & 
\multicolumn{1}{c}{Fraction decaying} \\ 
\multicolumn{1}{c}{Nucleus} &
\multicolumn{1}{c}{model} &
\multicolumn{1}{c}{$BR(\%)$} & 
\multicolumn{1}{c}{via $2^+_1$ state} \\
\hline
& & & \\[-3mm]
$^{62}$Ga &  MSDI & 0.28  &  80\%  \\
$^{66}$As &  MSDI & 0.67  &  70\%  \\
$^{70}$Br &  MSDI & 1.59  &  63\%  \\
$^{74}$Rb &  MSDI & 0.72  &  47\%  \\
          &  MSDI$^{\prime}$ & 0.92  &  56\%\footnotemark[1]  \\
\hline
\hline
\end{tabular}
\footnotetext[1]{We have taken the excited $0^+_2$ state to de-excite
44\% by electron-conversion to the ground state and
56\% to the first $2^+_1$ state as determined in ref. \protect\cite{Pi01}.}
\end{center}
\end{table}

}

These results clearly indicate that if superallowed $\beta$-decay $ft$-values
are to be determined for $A \geq 62$ nuclei with a precision better than, say,
$0.5 \%$, then new techniques will have to be developed to incorporate the effects
of many weak Gamow-Teller transitions.  Total absorption spectrometry has the
potential to accomplish this goal, but whether it can do so with sufficient precision
is an unanswered question.  For these heavy nuclei to become useful in testing $\delta_C$
calculations, the development of such new techniques will have to become a priority.

\acknowledgments

The work of JCH was supported by the U. S. Dept. of Energy under Grant
DE-FG03-93ER40773 and by the Robert A. Welch Foundation.
IST would like to thank the Cyclotron Institute of Texas A \& M University
for its hospitality during his two-month visit.

\end{document}